# Electronic Structures of Fe$_{3-x}$V$_x$Si Probed by Photoemission Spectroscopy[**]

Y. T. Cui[*1], A. Kimura[1], K. Miyamoto[1], K. Sakamoto[1], T. Xie[2], S. Qiao[2], M. Nakatake[2], K. Shimada[2], M. Taniguchi[1, 2], S. -i. Fujimori[3], Y. Saitoh[3], K. Kobayashi[3, 4], T. Kanomata[5], O. Nashima[5]

[1] Graduate School of Science, Hiroshima University, Higashi-Hiroshima, 739-8526, Japan
[2] Hiroshima Synchrotron Radiation Center, Hiroshima University, Higashi-Hiroshima 739-0046, Japan
[3] Japan Atomic Energy Agency (SPring-8/JAEA), Sayo, Hyogo 679-5198, Japan
[4] Japan Synchrotron Radiation Research Institute (SPring-8/JAERI), Sayo, Hyogo 679-5198, Japan
[5] Department of Applied Physics, Tohoku Gakuin University, Tagajo 985-8537, Japan

The electronic structures of the Heusler type compounds Fe$_{3-x}$V$_x$Si in the concentration range between $x = 0$ and $x = 1$ have been probed by photoemission spectroscopy (PES). The observed shift of Si 2$p$ core-level and the main valence band structres indicate a chemical potential shift to higher energy with increasing $x$. It is also clarified that the density of state at Fermi edge is owing to the collaboration of V 3$d$ and Fe 3$d$ derived states. Besides the decrease of the spectral intensity near Fermi edge with increasing $x$ suggests the formation of pseudo gap at large $x$.

## 1 Introduction

Heusler-type ternary and pseudobinary compounds $Y_2ZX$ ($Y$ and $Z$ denote transition metal atoms and $X$ is a metalloid) have received renewed interest owing to their ferromagnetic shape memory, magneto-resistance properties and thermoelectric power of novel mechanism etc. Among them, Fe$_{3-x}$V$_x$Si ($0 \leq x \leq 1$) alloys display a rich variety of physical behavior in electronic, magnetic and transport properties. The ordered Fe$_3$Si alloy forms D0$_3$-type structure (see Fig.1). In the unit cell, there are two in-equivalent Fe sites with specific neighbor configurations named as Fe$_I$ and Fe$_{II}$. The Fe$_I$ or Si sites are surrounded by eight nearest Fe$_{II}$ atoms in an octahedral configuration, while Fe$_{II}$ site has four Si and four Fe$_I$ nearest neighbors in a tetrahedral configuration. Then, Fe$_3$Si can be expressed as [Fe$_I$][Fe$_{II}$]$_2$Si. Since V lies to the left of Fe in the periodic table of elements, V atoms prefer to occupy Fe$_I$ sites [1]. With increasing vanadium concentration, the crystal structure of Fe$_{3-x}$V$_x$Si changes from D0$_3$ to L2$_1$ in the concentration range of $0 \leq x \leq 1$.

Fe$_{3-x}$V$_x$Si alloys are considered to have relatively high thermoelectric power (TEP) due to a pseudo gap near the Fermi level. Investigation performed by Nishino et al. [2] supports that Fe$_{3-x}$V$_x$Si compounds show anomalous resistivity properties including resistance maximum near the Curie point ($T_C$) and a negative slope of the resistivity above $T_C$, which increases with adding V atoms. Nashima et al. [3] reported that the compounds show a negative Seebeck coefficient at larger $x$, which indicates a semimetallic electronic state. The concentration dependence of the Seebeck coefficient on the V content suggests

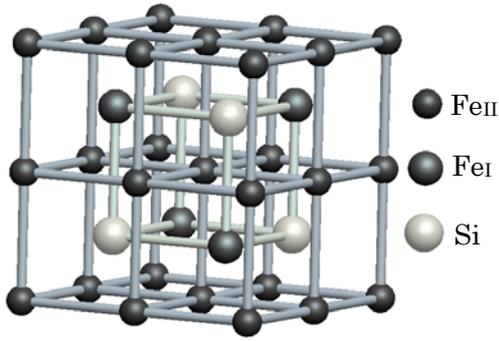

**Fig.1** Crystal structure of ordered $Fe_3Si$

- $Fe_{II}$
- $Fe_I$
- Si

that the substitution of Fe for V in $Fe_3Si$ causes not only the development of the pseudo gap but also a chemical potential shift to higher energy, which are also supported by the theoretical calculations [4, 5]. All of above mentioned physical behaviours could be mainly controlled by the electronic states close to Fermi level ($E_F$). So, in order to clarify those physical properties of $Fe_{3-x}V_xSi$, it is important to study their electronic structures directly by a photoemission spectroscopy (PES).

In this paper, the electronic structure in the valence band region will be investigated with synchrotron radiation in the vacuum ultraviolet and soft x-ray regions and a core level photoemission spectroscopy is also used to illustrate the chemical potential shift against the V concentration.

## 2 Experimental

$Fe_{3-x}V_xSi$ polycrystalline samples were prepared by repeated arc-melting of mixtures of Fe, V and Si under purified argon atmosphere. The subsequent post-annealing was performed at 1123 K for 5 days to improve the homogeneity of the alloys. Specimens were cut from the ingots by spark cutting with Cu wire to a size of $1.2 \times 1.2 \times 5$ $mm^3$.

The experiment was carried out with synchrotron radiation in the vacuum ultraviolet (VUV) and soft X-ray regions. The VUV photoemission spectroscopy (VUV-PES) was measured with a hemispherical analyzer at BL-7 of Hiroshima Synchrotron Radiation Center (HSRC). The soft X-ray photoelectron spectroscopy (SX-PES) was also measured at the undulator beamline BL-23SU of SPring-8. The angle between the incident photon beam and the lens axis of the analyzer was 45˚. The acceptance angle of the analyzer was set to ±8˚. The Fermi energy and the total energy resolution were checked with the evaporated Au film. The evaluated total energy resolutions for the VUV photoemission at 50 K were 54 meV and 190 meV for the photon energies ($hv$) of 38 eV and 146 eV, respectively. For the soft X-ray photoemission at room temperature the total energy resolution was 300 meV for $hv$ = 1030 eV.

Clean surfaces of the specimens were obtained by in situ fracturing with a knife edge, and they were confirmed by the O $1s$ core level for the SX-PES as well as the O $2p$ signal in the valence band for VUV-PES.

## 3 Result and discussion

The Si $2p$ core level spectra have been measured at $hv$ = 146 eV (see Fig.2). We plot the binding energies of $2p_{3/2}$ and $2p_{1/2}$ main peaks as a function of $x$ in Fig.3. It is clearly found that the Si $2p$ main peaks shift to higher binding energy with increasing $x$. The observed core-level shift is possibly due to a chemical potential shift to higher energy. There would also be another possibility of a chemical shift due to a different environment of Si atom with different vanadium concentration. However, since the V atom prefers to occupy the $Fe_I$ site, the Si atom is still surrounded by 8 $Fe_{II}$ atoms while keeping the environment of Si, then the chemical shift should be excluded. Besides the Si $2p$ main peaks, there is a somewhat broad structure at lower binding energy, it is also recognized that the Si $2p$ main peaks are broader for the larger $x$, which might come from an atomic disorder at the surface or inside the crystal. In order to find the origin of these additional structures, the Si 2$p$ core-level spectra have been measured again using soft x-ray ($h\nu = 1030$ eV) at room temperature (for $x = 0$ and 0.8) and are plotted in the inset of Fig.2. Note that the additional structures at low binding energy are markedly reduced in the SX-PES spectra. This result indicates that they are mainly due to the surface effect if we take into account the shorter electron mean-free path for the electrons with smaller kinetic energy.

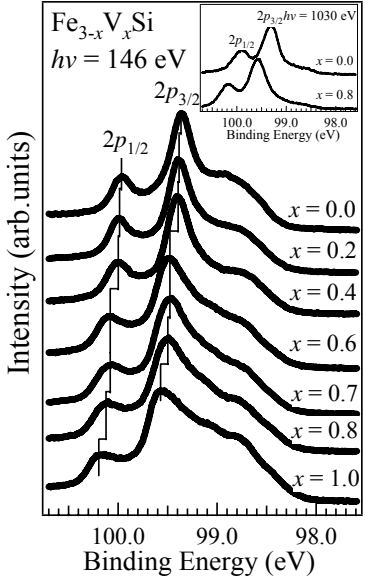
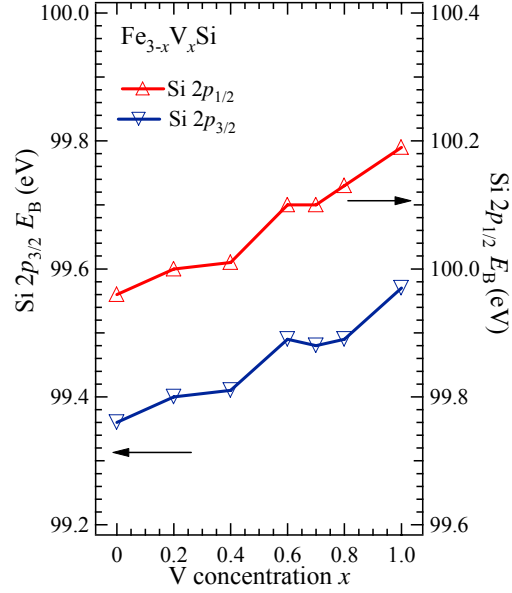

**Fig.2** Si 2$p$ core-level photoemission spectra of Fe$_{3-x}$V$_x$Si with $x = 0\sim 1$ measured at $h\nu = 146$ eV.

**Fig.3** Binding energies of Si 2$p_{3/2}$ and 2$p_{1/2}$ main peaks vs. $x$.

The valence band spectra measured at $h\nu = 38$ eV and $T = 50$ K with different $x$ are shown in Fig.4. We find two sharp peaks denoted as A (just below $E_F$) and C (at 1.5 eV below $E_F$) for $x = 0$. It is noticed that the structure C shifts to higher binding energy ($E_B$) and is less weighted with increasing $x$. Besides, the intensity of peak A is also gradually reduced and the peak position does not change toward larger $x$. For $x > 0.4$, the peak B develops around 1 eV and it becomes sharper and shifts to larger $E_B$ with increasing $x$. In order to assign these main structures, the valence band spectra taken at the different photon energies for $x = 0$ and 0.8 are compared in Fig.5. It implies that the intensities at $E_F$ are enhanced in the VUV-PES spectra for both $x = 0$ and 0.8 compared to those in the SX-PES spectra, which are more bulk sensitive. In Fig.5, the photoemission intensities for the structures E and F are enhanced at the higher photon energy.

By taking into account the photoionization cross section, some structures can be separated into different component contributions, because the cross section of V 3$d$ state decreases faster than that of Fe 3$d$ state, while that of Si 3$p$ state decreases more moderately than that of Fe 3$d$ state with increasing photon energy. It is considered that the structures E and F should be Si 3$p$ and Si 3$s$ derived states, respectively. The spectral intensity at $E_F$ for $x = 0.8$ is much reduced with increasing photon energy compared to that for $x = 0$, which means the density of states at $E_F$ should be mainly owing to the contribution of V 3$d$ state for large $x$. The structure B emerges and enhanced with adding V atoms at $h\nu = 38$ eV, which should be related to the V 3$d$ states. While for $x = 0.8$, with increasing $h\nu$, the spectral intensity of the structure B does not change so much as that at $E_F$, which means it is mainly due to the contribution of Fe 3$d$ states. Although it seems difficult for the structure D to be assigned only from the cross section, this may be due to a bonding state of Si $3p$ orbital with Fe $3d$ or V $3d$ orbitals. The observed shift of the structures B and C to higher binding energy is another evidence for the chemical potential shift. In Fig.6, it can also be seen that the intensity of the spectra below $E_F$ is reduced with increasing V concentration especially for $x > 0.6$. That might be related to the formation of the pseudo gap at large $x$ as pointed out by the theoretical calculation [4, 5] and the experiment [3].

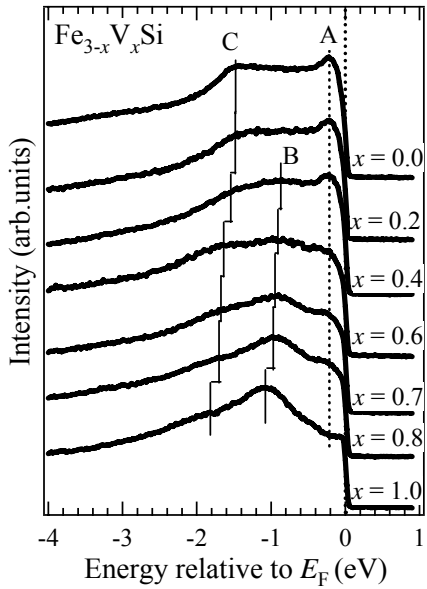

**Fig.4** Valence band photoemission spectra of $Fe_{3-x}V_xSi$ with photon energy ($h\nu$) of 38 eV at 50 K. The black dotted lines are just for the guide to the eye.

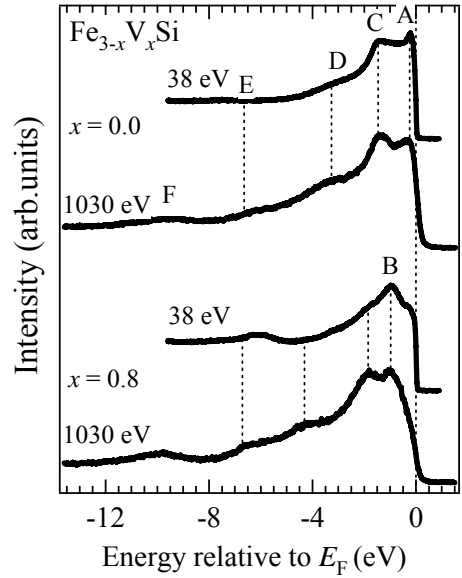

**Fig.5** Valence band photoemission spectra of $Fe_{3-x}V_xSi$ with $x = 0.0$ and $0.8$ measured at $h\nu = 38$ eV and 1030 eV.

## 4 Conclusion

In conclusion, the valence band and Si $2p$ core-level photoemission spectra of $Fe_{3-x}V_xSi$ ($0 \leq x \leq 1$) have been measured with synchrotron radiation in the vacuum ultraviolet and soft X-ray regions. The observed energy shift of Si $2p$ core-level and the main valence band structures to higher binding energy with increasing $x$ is due to the chemical potential shift to higher energy with increasing V concentration. The features observed at 1 eV and 1.5~1.8 eV for several $x$ have been assigned to the Fe $3d$ derived state by taking into account the photoionization cross section. It has also been clarified that the spectral intensity at the Fermi level consists of both V $3d$ and Fe $3d$ derived states. The decrease of the spectral intensity near $E_F$ with increasing $x$ suggests the formation of pseudo gap at large $x$.

**Acknowledgements**    This work was done under the approval of the Spring-8 Proposal Assessing Committee (Proposal No. 2005B3811). The work was supported by the Ministry of Education, Culture, Sports, Sciences and Technology. The authors also would like to thank Professor Soda for the helpful discussion.